\documentclass[twocolumn,preprintnumbers,prb,aps,amssymb,showpacs,superscriptaddress,nobibnotes]{revtex4}

\usepackage{graphicx}
\usepackage{natbib}
\usepackage{dcolumn}
\usepackage{bm} 
\usepackage{color}
\usepackage{setspace}
\usepackage[colorlinks=true,linkcolor=blue,urlcolor=blue,citecolor=blue]{hyperref}

\begin{document}

\newcommand{\lsco}{La$_{2-x}$Sr$_x$CuO$_4$}
\newcommand{\lbco}{La$_{2-x}$Ba$_x$CuO$_4$}
\newcommand{\lnsco}{La$_{1.6-x}$Nd$_{0.4}$Sr$_x$CuO$_4$}
\newcommand{\lesco}{La$_{1.8-x}$Eu$_{0.2}$Sr$_x$CuO$_4$}
\newcommand{\lresco}{La$_{2-x-y}$RE$_{y}$Sr$_x$CuO$_4$}

\newcommand{\lnscoxy}{La$_{2-x-y}$Nd$_{y}$Sr$_x$CuO$_4$}
\newcommand{\lnco}{La$_{1.6}$Nd$_{0.4}$CuO$_4$}
\newcommand{\pcco}{Pr$_{2-x}$Ce$_{x}$CuO$_{4}$}
\newcommand{\ybco}{YBa$_{2}$Cu$_{3}$O$_{y}$}
\newcommand{\tltwotwoone}{Tl$_{2}$Ba$_{2}$CuO$_{6+\delta}$}
\newcommand{\hbco}{HgBa$_{2}$CuO$_{4+\delta}$}
\newcommand{\bsco}{Bi$_2$Sr$_{2}$CuO$_{6+\delta}$}
\newcommand{\bscco}{Bi$_2$Sr$_{2}$CaCu$_{2}$O$_{8+\delta}$}

\newcommand{\muohmcm}[1]{#1\,$\mu\Omega$\,cm}

\newcommand{\TN}{$T_{\rm N}$}
\newcommand{\Tc}{$T_{\rm c}$}
\newcommand{\Tstar}{$T^{\star}$}
\newcommand{\TCDW}{$T_{\rm CDW}$}
\newcommand{\Hc}{$H_{\rm c2}$}

\newcommand{\ie}{{\it i.e.}}
\newcommand{\eg}{{\it e.g.}}
\newcommand{\etal}{{\it et al.}}

%%%%%%%%%%%%%%%%%%%%%%%%%%%% TITLE

\title{Two types of nematicity in the phase diagram of the cuprate superconductor \ybco}

%%%%%%%%%%%%%%%%%%%%%%%%%%%% AUTHORS

\author{O. Cyr-Choini\`{e}re}
\affiliation{D\'{e}partement de physique  \&  RQMP, Universit\'{e} de Sherbrooke, Sherbrooke,  Qu\'{e}bec J1K 2R1, Canada}

\author{G. Grissonnanche}
\affiliation{D\'{e}partement de physique  \&  RQMP, Universit\'{e} de Sherbrooke, Sherbrooke,  Qu\'{e}bec J1K 2R1, Canada}

\author{S. Badoux}
\affiliation{D\'{e}partement de physique  \&  RQMP, Universit\'{e} de Sherbrooke, Sherbrooke,  Qu\'{e}bec J1K 2R1, Canada}

\author{J. Day}
\affiliation{Department of Physics and Astronomy, University of British Columbia, Vancouver, British Columbia V6T 1Z4, Canada}

\author{D.~A.~Bonn}
\affiliation{Department of Physics and Astronomy, University of British Columbia, Vancouver, British Columbia V6T 1Z4, Canada}
\affiliation{Canadian Institute for Advanced Research, Toronto, Ontario M5G 1Z8, Canada}

\author{W. N. Hardy}
\affiliation{Department of Physics and Astronomy, University of British Columbia, Vancouver, British Columbia V6T 1Z4, Canada}
\affiliation{Canadian Institute for Advanced Research, Toronto, Ontario M5G 1Z8, Canada}

\author{R. Liang}
\affiliation{Department of Physics and Astronomy, University of British Columbia, Vancouver, British Columbia V6T 1Z4, Canada}
\affiliation{Canadian Institute for Advanced Research, Toronto, Ontario M5G 1Z8, Canada}

\author{N. Doiron-Leyraud}
\affiliation{D\'{e}partement de physique  \&  RQMP, Universit\'{e} de Sherbrooke, Sherbrooke,  Qu\'{e}bec J1K 2R1, Canada}

\author{Louis~Taillefer}
\email{louis.taillefer@usherbrooke.ca}
%\altaffiliation{Correspondence should be addressed to L.T. (e-mail: louis.taillefer@usherbrooke.ca)}
\affiliation{D\'{e}partement de physique  \&  RQMP, Universit\'{e} de Sherbrooke, Sherbrooke,  Qu\'{e}bec J1K 2R1, Canada}
\affiliation{Canadian Institute for Advanced Research, Toronto, Ontario M5G 1Z8, Canada}

\date{\today}

%%%%%%%%%%%%%%%%%%%%%%%%%%%%%%%%%%%%%%%%%%%%%
%%%%%%%%%%%%%                  ABSTRACT                     %%%%%%%%%%%%%%
%%%%%%%%%%%%%%%%%%%%%%%%%%%%%%%%%%%%%%%%%%%%%

\begin{abstract}

Nematicity has emerged as a key feature of cuprate superconductors, 
but its link to other fundamental properties such as superconductivity, charge order and the pseudogap remains unclear. 
Here we use measurements of transport anisotropy in \ybco~to distinguish two types of nematicity. 
The first is associated with short-range charge-density-wave modulations in a doping region near $p = 0.12$. 
It is detected in the Nernst coefficient, but not in the resistivity. 
The second type prevails at lower doping, where there are spin modulations but no charge modulations. 
In this case, the onset of in-plane anisotropy $-$ detected in both the Nernst coefficient and the resistivity $-$ 
follows a line in the temperature-doping phase diagram that tracks the pseudogap energy. 
We discuss two possible scenarios for the latter nematicity.
%The topology of the resulting phase diagram suggests that a pseudogap phase with nematic character 
%shapes the domes of both superconductivity and charge order.

\end{abstract}

%%%%%%%%%%%%%%%%%%%%%%%%%%%%%%%%%%%%%%%%%%%%%%%%%%%%%%%%%%%%%%%%%%%%%%%%%


% 74.72.Gh	Hole-doped cuprate superconductors
% 74.25.Dw	Phase diagrams superconductivity
% 74.25.F-	Transport properties

%DOI: 10.1103/PhysRevB.00.004500

\maketitle

%%%%%%%%%%%%%%%%%%%%%%%%%%%%%%%%%%%%%%%%%%%%%
%%%%%%%%%%%                  INTRODUCTION                     %%%%%%%%%%%%%
%%%%%%%%%%%%%%%%%%%%%%%%%%%%%%%%%%%%%%%%%%%%%

\section{INTRODUCTION}

\begin{figure}[b!]
\centering
\includegraphics[width=0.46\textwidth]{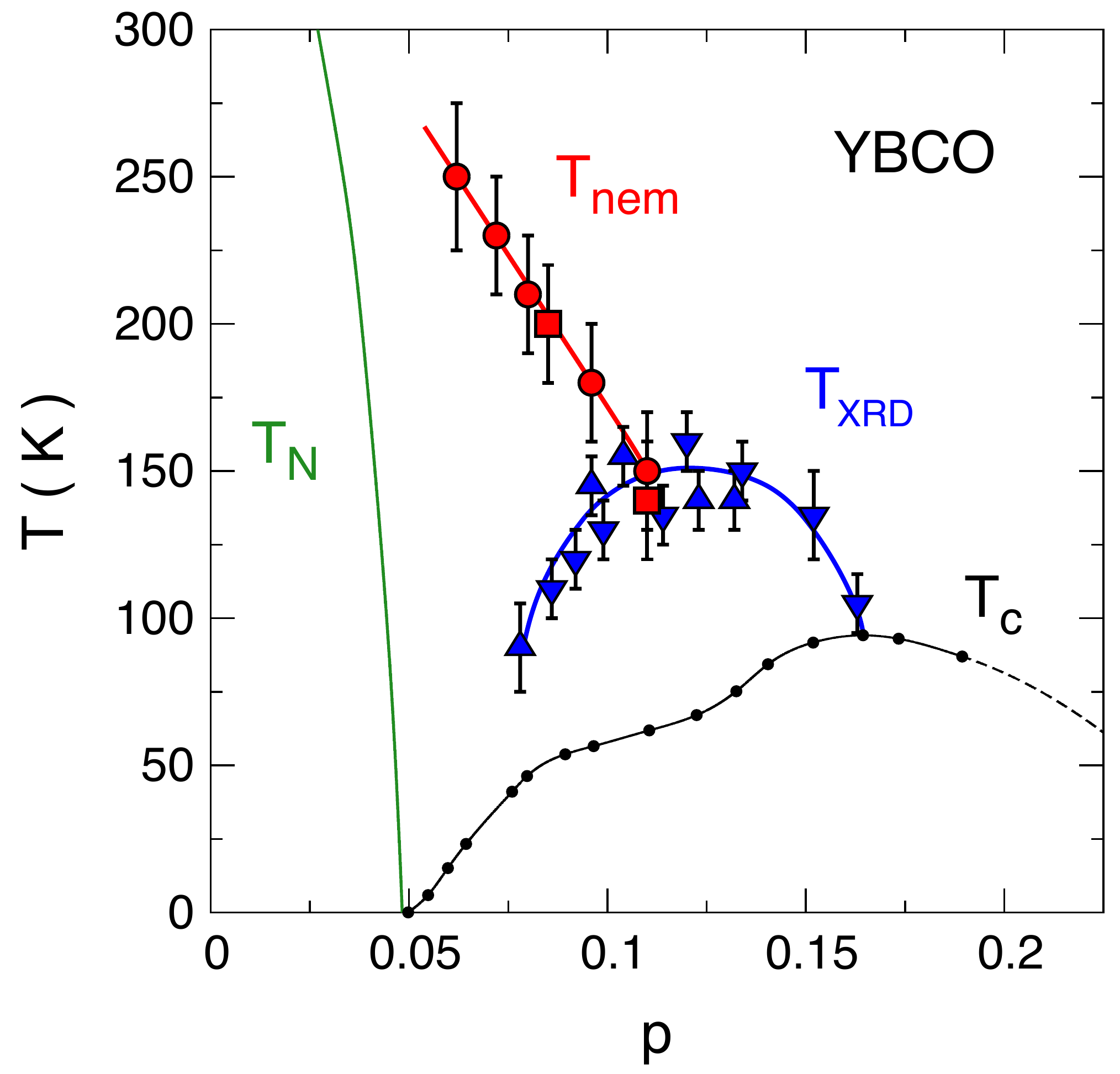}
\caption{Temperature-doping phase diagram of YBCO, with the superconducting phase below 
	$T_{\rm c}$ (black dots~[\onlinecite{Liang2006}]) and the antiferromagnetic phase below $T_{\rm N}$ (green line). 
	Short-range charge-density-wave (CDW) modulations are detected 
	% by X-ray diffraction 
	below $T_{\rm XRD}$ (up triangles;~[\onlinecite{Hucker2014}] down triangles~[\onlinecite{Blanco-Canosa2014}]). 
	%The temperature 
	$T_{\rm nem}$ (red symbols) marks the onset of the rise in the resistivity anisotropy 
	(circles; from ref.~\onlinecite{Ando2002} and Fig.~\ref{Anirho-AniNernst}(b)) 
	and in the Nernst anisotropy (squares; from Figs.~\ref{Anirho-AniNernst}(c) and \ref{Anirho-AniNernst}(d)), as temperature is reduced. 
	%Error bars indicate the uncertainty in defining the various onset temperatures. 
	The blue and red lines are guides to the eye. 
	They delineate the two regions of the phase diagram where two types of nematicity prevail 
	%(see text).
	$-$ one associated with CDW modulations (blue), the other not (red). 
	}
\label{Phasediagram}
\end{figure}

It has become clear that charge-density-wave (CDW) order is a generic tendency of cuprate 
superconductors~[\onlinecite{Wu2011,Ghiringhelli2012,Chang2012a,Croft2014,Tabis2014,Comin2014,daSilvaNeto2014}].
In \ybco~(YBCO), short-range CDW modulations detected by X-ray diffraction (XRD) appear below a doping 
$p \sim 0.16$~[\onlinecite{Hucker2014,Blanco-Canosa2014}]. 
Both the amplitude of the CDW modulations and their onset temperature, $T_{\rm XRD}$, peak at 
$p = 0.12$~(Fig.~\ref{Phasediagram})~[\onlinecite{Hucker2014,Blanco-Canosa2014}]. 
NMR measurements in high magnetic fields~[\onlinecite{Wu2011,Wu2013}] detect the abrupt onset of CDW order below a temperature $T_{\rm NMR}$ that also peaks at $p = 0.12$. 
This CDW order causes a reconstruction of the Fermi surface~[\onlinecite{Taillefer2009}] 
detected as a sign change in the Hall~[\onlinecite{LeBoeuf2007,LeBoeuf2011}] 
and Seebeck~[\onlinecite{Chang2010,Laliberte2011}] coefficients, from positive at high temperature to negative at low temperature, 
due to the formation of a small electron-like pocket~[\onlinecite{Doiron-Leyraud2007}]. 
The temperature $T_{\rm max}$ below which $R_{\rm H}(T)$ drops as a result of Fermi-surface reconstruction peaks at $p = 0.12$~[\onlinecite{LeBoeuf2011}]. 
The CDW order and associated reconstruction, both peaked at $p = 0.12$, cause a dip in $T_{\rm c}$~(ref.~\onlinecite{Liang2006}) (Fig.~\ref{Phasediagram}) 
and a local minimum in the superconducting upper critical field $H_{\rm c2}$~(ref.~\onlinecite{Grissonnanche2014}) at $p = 0.12$.

%%%%%%%%%%%%%%%%%   Figure 1 Phase diagram  %%%%%%%%%%%%%%%%%%%%%%%%%%%%%%%%%%%%%%%%%%%%%%%%%

%%%%%%%%%%%%%%%%%%%%%%%%%%%%%%%%%%%%%%%%%%%%%%%%%%%%%%%%%%%%%%%%%%%%%%%

\begin{figure}
\centering
\includegraphics[width=0.46\textwidth]{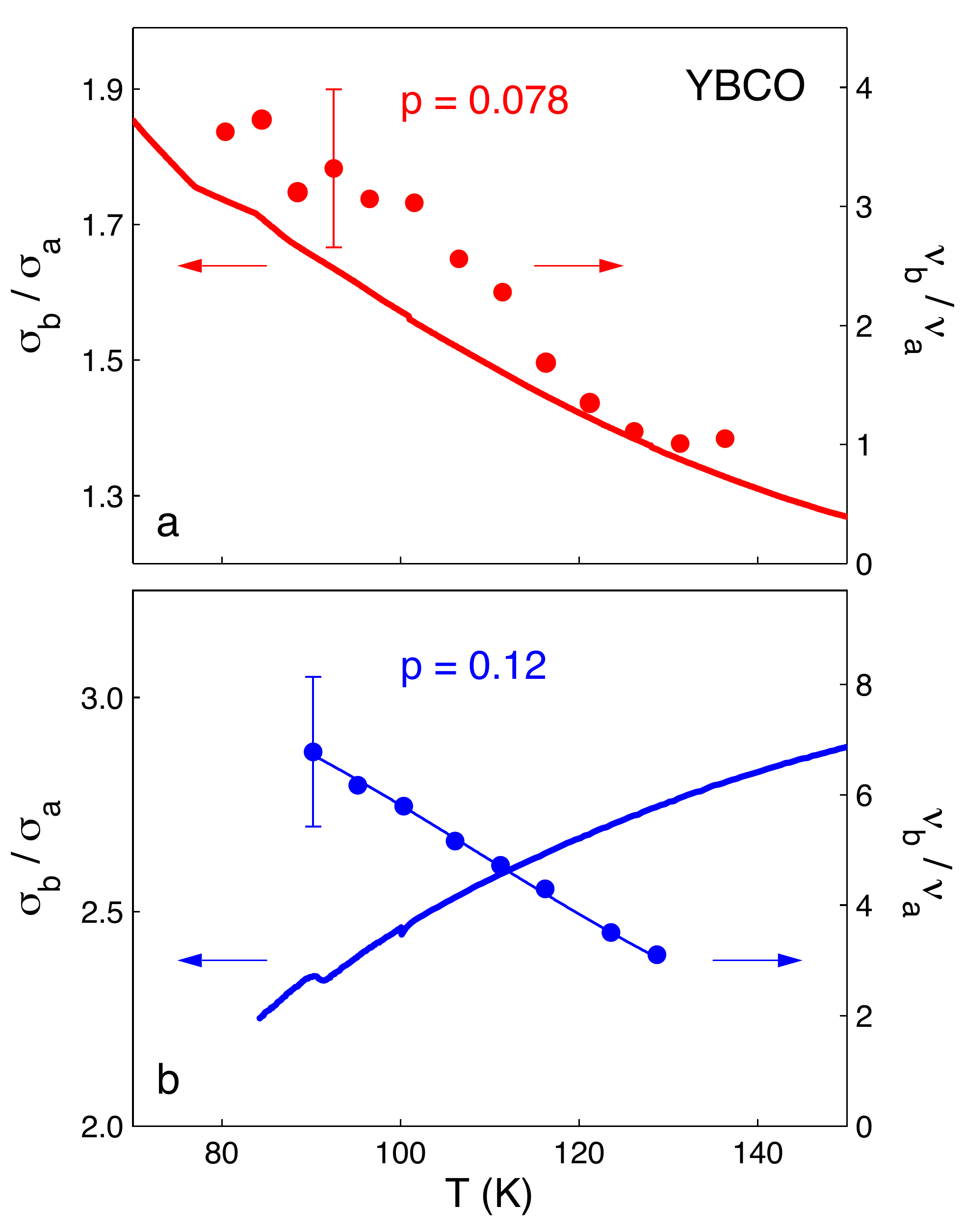}
\caption{Anisotropy of the Nernst coefficient $\nu(T)$ compared with the corresponding anisotropy of the conductivity $\sigma(T)$ in YBCO, 
	both plotted as ratios: $\nu_b\,/\,\nu_a$ (dots) and $\sigma_b\,/\,\sigma_a$ (curve) (adapted from ref.~\onlinecite{Daou2010}). 
	The CuO chains that run along the $b$ axis of the orthorhombic structure make a contribution to $\sigma_b$ such that
	$\sigma_b\,/\,\sigma_a > 1$ at $T = 300$~K and $\sigma_b\,/\,\sigma_a$ decreases below $T = 150$~K~[\onlinecite{Ando2002}]. 
	(a) At $p = 0.078$, both $\nu_b\,/\,\nu_a$ (dots) and $\sigma_b\,/\,\sigma_a$ rise with decreasing temperature, so that	
	both reveal the emergence of nematicity in the CuO$_2$ planes at low temperature. 
	(b) At $p = 0.12$, nematicity is observed in $\nu_b\,/\,\nu_a$, as a huge anisotropy that rises with decreasing $T$. 
	However, no corresponding anisotropy rise is detected in $\sigma_b\,/\,\sigma_a$, which only shows a drop typical of that due to the CuO chains~[\onlinecite{Ando2002}]. 
	Those two comparisons therefore reveal an empirical difference in the nematicities at $p = 0.078$ and
	at $p = 0.12$. 
	}
\label{Anirho-Aninu-p08-p12}
\end{figure}

\begin{figure}
\centering
\includegraphics[width=0.46\textwidth]{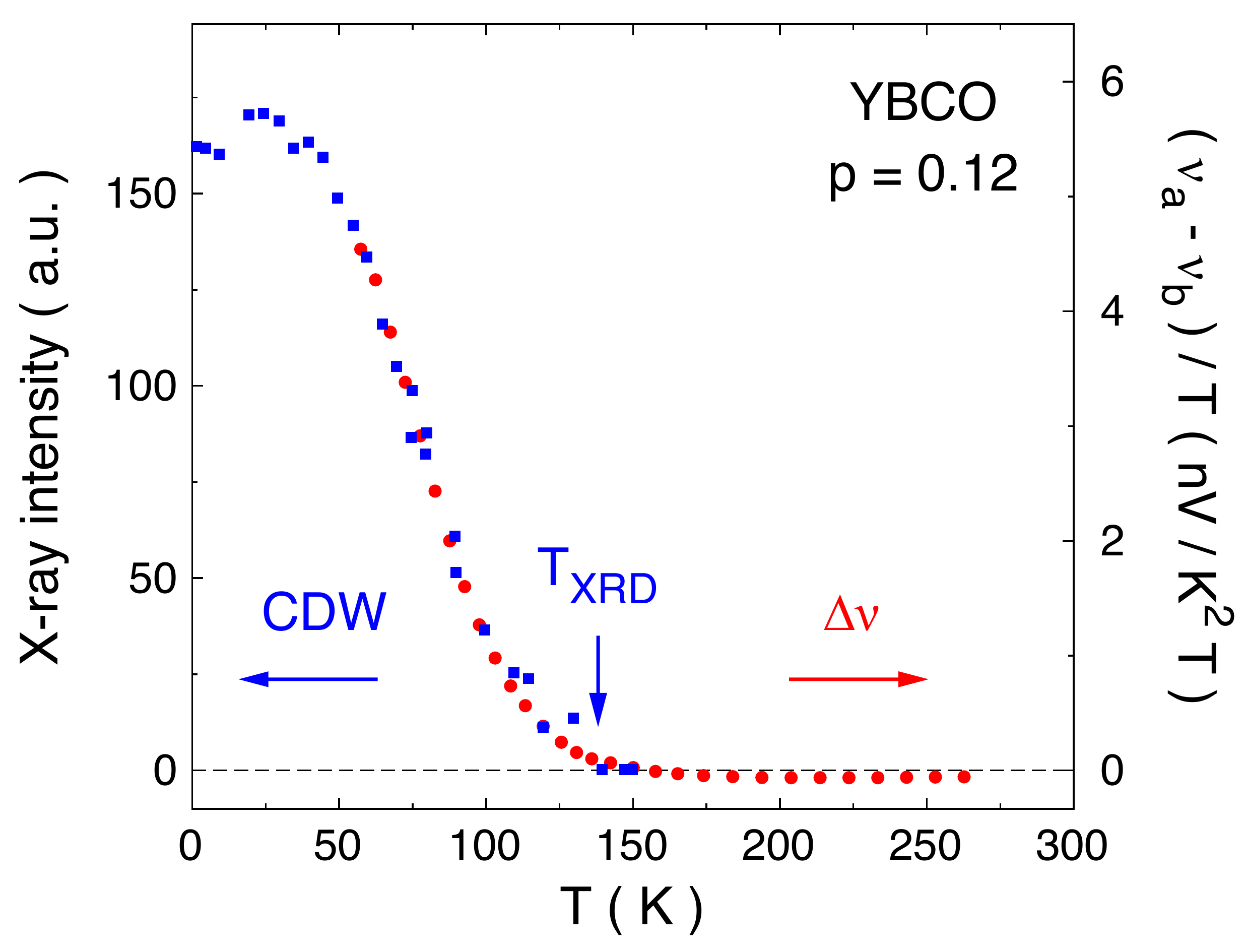}
\caption{Temperature dependence of the X-ray intensity associated with the short-range CDW modulations in YBCO 
	at a doping $p = 0.12$ and a magnetic field $H = 15$\,T (blue squares, left axis; from ref.~\onlinecite{Chang2012a}). 
	Note that above $T_{\rm c} = 66$\,K the data at $H = 0$ are identical to the data at $H = 15$\,T~[\onlinecite{Chang2012a}]. 
	$T_{\rm XRD} = 140$\,K marks the approximate onset of the gradual rise in CDW intensity (blue arrow). 
	The in-plane anisotropy of the Nernst coefficient $\nu_x(T)$ measured at the same doping and the same field 
	is plotted as $(\,\nu_a - \nu_b\,)\, /\,T$ (red dots, right axis; from ref.~\onlinecite{Daou2010}). 
	It is seen to track the CDW intensity very well, compelling evidence that the anisotropic CDW modulations cause the large Nernst anisotropy at that doping. 
	At dopings above $p = 0.12$, the rise in the Nernst anisotropy continues to track the onset of CDW modulations, 
	\textit{i.e.} the different curves of $(\,\nu_a - \nu_b\,)\, /\,T$ for different dopings scale when plotted as $T\,/\,T_{\rm XRD}$ (see Fig.~\ref{Dnu-Tnu-TXRD}(b)). 
	}
\label{Dnu-RX}
\end{figure}

The short-range CDW modulations are known to be anisotropic within the CuO$_2$ planes of YBCO,~[\onlinecite{Hucker2014,Blanco-Canosa2014}] 
and they appear to break the rotational (C4) symmetry of the planes,~[\onlinecite{Wu2015,Comin2015}] 
\textit{i.e.} they have a nematic character. 
This should give rise to an in-plane anisotropy of transport. 
A large in-plane anisotropy of the Nernst coefficient $\nu(T)$ has indeed been observed in YBCO (see Fig.~\ref{Anirho-Aninu-p08-p12})~[\onlinecite{Daou2010}], 
and it does grow in tandem with the CDW modulations measured by XRD at $p = 0.12$ (see Fig.~\ref{Dnu-RX}).

%%%%%%%%%%%%%%%%%   Figure Anirho-Aninu-p08-p12  %%%%%%%%%%%%%%%%%%%%%%%%%%%%%%%%%%%%%%%%%%%%%%%%%

%%%%%%%%%%%%%%%%%   Figure  Dnu vs RX   %%%%%%%%%%%%%%%%%%%%%%%%%%%%%%%%%%%%%%%%%%%%%%%%%

%%%%%%%%%%%%%%%%%%%%%%%%%%%%%%%%%%%%%%%%%%%%%%%%%%%%%%%%%%%%%%%%%%%%%%%

\begin{figure*}
\centering
\includegraphics[width=0.85\textwidth]{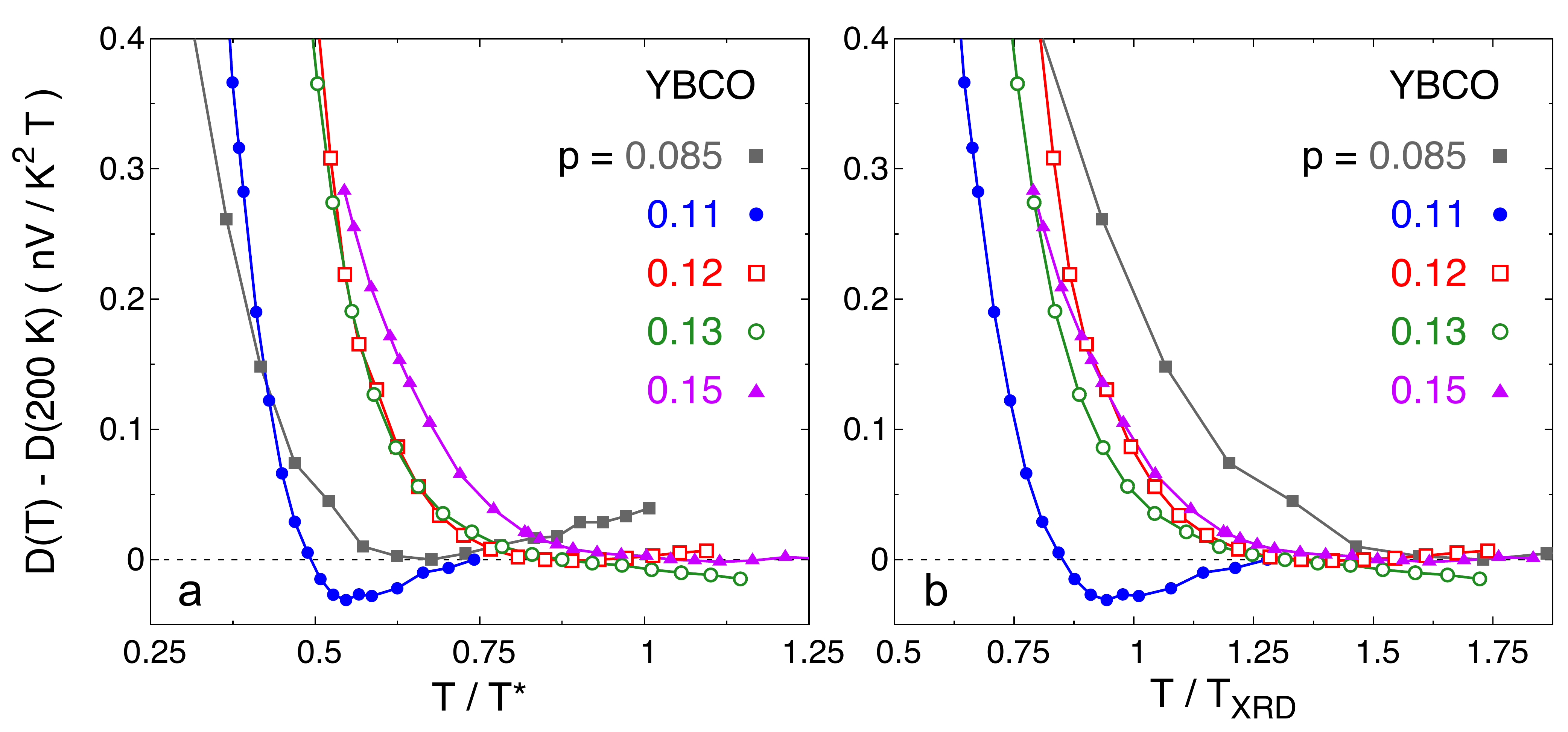}
\caption{Anisotropy difference in the Nernst coefficient of YBCO, defined as $D(T) = (\,\nu_a - \nu_b\,)\, /\,T$, 
	and plotted as $D(T) - D(200~{\rm K})$ vs $T\,/\,T^{\star}$ in panel (a), and as $D(T) - D(200~{\rm K})$ vs $T\,/\,T_{\rm XRD}$ in panel (b), 
	where $T^{\star}$ is given by the dashed line in Fig.~\ref{Ando-Curvature-map}	and $T_{\rm XRD}$ by the blue line in Fig.~\ref{Phasediagram}. 
	Data at $p = 0.12, 0.13$ and $0.15$ are from ref.~\onlinecite{Daou2010}. 
	Panel (a) shows that the anisotropy at low temperature is not controlled by the pseudogap temperature $T^{\star}$ across all dopings. 
	In panel (b), the collapse of the three curves at $p = 0.12, 0.13$ and $0.15$ onto a common curve shows that at those dopings 
	the upturn in $D(T)$ at low $T$ is correlated with the onset of CDW modulations (at $T_{\rm XRD}$), in agreement with Fig.~\ref{Dnu-RX}. 
	By contrast, at $p = 0.085$ and $0.11$, the anisotropy is not linked to the CDW order. 
	Indeed, those two curves at lower doping are correlated instead with $T_{\rm nem}$, the onset of anisotropy in the resistivity (Fig.~\ref{Phasediagram}). 
	}
\label{Dnu-Tnu-TXRD}
\end{figure*}

In the original report~[\onlinecite{Daou2010}], for dopings $p = 0.12$ and higher ($0.13, 0.15$ and $0.18$) the Nernst anisotropy difference 
was shown to collapse onto a common curve when plotted as $(\,\nu_a - \nu_b\,)\,/\,T$ vs $T\,/\,T^{\star}$ (Fig. 4 of ref.~\onlinecite{Daou2010}), 
where $T^{\star}$ is the pseudogap temperature defined as the onset of the drop in $\nu_{x}\,/\,T$ vs $T$ below its constant value at high temperature 
(inset in Fig.~1 of ref.~\onlinecite{Daou2010}). 
(This definition was shown to be equivalent, within error bars, to the usual definition of $T^{\star}$ 
from the downward deviation in the resistivity $\rho_a(T)$ vs $T$ from its linear dependence at high temperature (Fig. S4 of ref.~\onlinecite{Daou2010}).)
What this collapse revealed is a common, very slow increase of Nernst anisotropy starting at $T^{\star}$, for $p = 0.12$ and higher (see Fig.~\ref{Dnu-Tnu-TXRD}(a)).
This led to the conclusion that the pseudogap phase causes an extra anisotropy in transport. 

However, as shown in Fig.~\ref{Dnu-Tnu-TXRD}(b), the different curves scale just as well when plotted as $T\,/\,T_{\rm XRD}$, perhaps even better. 
Note, moreover, that the growth in $(\,\nu_a - \nu_b\,)\, /\,T$ that occurs between $T^{\star}$ and $T_{\rm XRD}$ is only 1\% 
of the growth that takes place below $T_{\rm XRD}$ , tracking the XRD intensity (Fig.~\ref{Dnu-RX}). 
Therefore, the large Nernst anisotropy in YBCO at $p = 0.12$ (and higher doping) is more likely to be caused by nematic CDW modulations. 
This is confirmed by looking at lower dopings (see below).

%%%%%%%%%%%%%%%%%   Figure  Dnu vs Tnu and TXRD  %%%%%%%%%%%%%%%%%%%%%%%%%%%%%%%%%%%%%%%%%%%%%%%%%

%%%%%%%%%%%%%%%%%%%%%%%%%%%%%%%%%%%%%%%%%%%%%%%%%%%%%%%%%%%%%%%%%%%%%%%

There is a second type of nematicity in YBCO, which is not associated with CDW modulations. 
The signature of this nematicity is an in-plane anisotropy in the resistivity $\rho(T)$ that is distinct from the anisotropy due to the CuO chains. 
It shows up as an upturn in the anisotropy ratio $\rho_a\,/\,\rho_b$ at low temperature, first detected at low doping ($p < 0.1$)~[\onlinecite{Ando2002}].
In particular, it is observed at dopings where there are no CDW modulations (Fig.~\ref{Phasediagram}) and no Fermi-surface reconstruction~[\onlinecite{LeBoeuf2011}], 
below $p = 0.08$.
In that region, the Nernst anisotropy, $\nu_b\,/\,\nu_a$, was found to track the conductivity anisotropy, 
$\sigma_b\,/\,\sigma_a\,(=\,\rho_a\,/\,\rho_b)$ ~[\onlinecite{Daou2010}] (see Fig.~\ref{Anirho-Aninu-p08-p12}(a)). 
This second kind of nematicity may be associated with spin degrees of freedom, since an in-plane anisotropy in the spin fluctuation spectrum 
measured by inelastic neutron scattering emerges spontaneously upon cooling~[\onlinecite{Hinkov2008}], 
in tandem with the rise in $\rho_a\,/\,\rho_b$~[\onlinecite{Ando2002}].

%%%%%%%%%%%%%%%%%%%%%%%%%%%%%%%%%%%%%%%%%%%%%%%%%%%%%%%%%%%%%%%%%%%%%%%

In this Article, we report measurements of $\rho(T)$ and $\nu(T)$ in YBCO performed
in the intermediate doping range $0.08 < p < 0.12$, and use the data to disentangle the CDW-related anisotropy
from the nematicity at low doping.
We define the onset temperature for the latter nematicity, $T_{\rm nem}$, and track it as a function of doping (Fig.~\ref{Phasediagram}).
We find that $T_{\rm nem}$ hits the CDW dome ($T_{\rm XRD}$) near its peak 
and that it scales with the pseudogap energy $E_{\rm PG}$.

%%%%%%%%%%%%%%%%%%%%%%%%%%%%%%%%%%%%%%%%%%%%%
%%%%%%%%%%%%%                  METHODS                     %%%%%%%%%%%%%%
%%%%%%%%%%%%%%%%%%%%%%%%%%%%%%%%%%%%%%%%%%%%%

\section{METHODS}
Single crystals of \ybco{} (YBCO) were prepared by flux growth~[\onlinecite{Liang2012}]. 
Their hole concentration (doping) $p$ is determined from the superconducting transition temperature $T_{\rm c}$~[\onlinecite{Liang2006}], 
defined as the temperature below which the zero-field resistance is zero.
A high degree of oxygen order was achieved for a pair of samples with $p = 0.11$ ($y = 6.54$, ortho-II order) 
and with $p = 0.12$ ($y = 6.67$, ortho-VIII order). 
Each measurement of anisotropy was done on a pair of samples oriented with the long transport direction 
along the $a$ and $b$ axis of the crystalline detwinned structure, respectively. 
The in-plane electrical resistivity of YBCO was measured at dopings $p = 0.11$ and $0.12$, in each case on a pair of $a$~axis and $b$~axis samples (Fig.~\ref{Anirho-AniNernst}(b)). 
The Nernst effect -- the transverse voltage (along $y$) generated by a longitudinal thermal current (along $x$) 
in the presence of a perpendicular magnetic field (along $z$) -- was measured as described elsewhere~[\onlinecite{Daou2010}]. 
The Nernst response of YBCO was measured at dopings $p = 0.085$ ($H = 16$\,T) and $p = 0.11$ ($H = 18$\,T), in each case on a pair of $a$~axis and $b$~axis samples 
(Figs.~\ref{Dnu-Tnu-TXRD},~\ref{Anirho-AniNernst}(c) and~\ref{Anirho-AniNernst}(d)). 
The thermal current was induced along the $a$~($b$)~axis and the voltage was measured along the transverse $b$~($a$)~axis for $a$~($b$)~axis samples. 
In all measurements, the magnetic field was applied along the $c$~axis, normal to the CuO$_2$ planes. 
For the temperature range of data shown in Fig.~\ref{Anirho-AniNernst}, the Nernst coefficient (and its anisotropy) is independent of magnetic field strength.

%%%%%%%%%%%%%%%%%%%%%%%%%%%%%%%%%%%%%%%%%%%%%%%%%%%%%%%%%%%%%%%%%%%%%%%

%%%%%%%%%%%%%%%%%%%%%%%%%%%%%%%%%%%%%%%%%%%%%
%%%%%%%%%%%%%                  RESULTS                     %%%%%%%%%%%%%%
%%%%%%%%%%%%%%%%%%%%%%%%%%%%%%%%%%%%%%%%%%%%%

\section{RESULTS}

%%%%%%%%%%%%%%%%%   Figure  Anirho-AniNernst  %%%%%%%%%%%%%%%%%%%%%%%%%%%%%%%%%%%%%%%%%%%%%%%%%

\begin{figure*}
\centering
\includegraphics[width=0.95\textwidth]{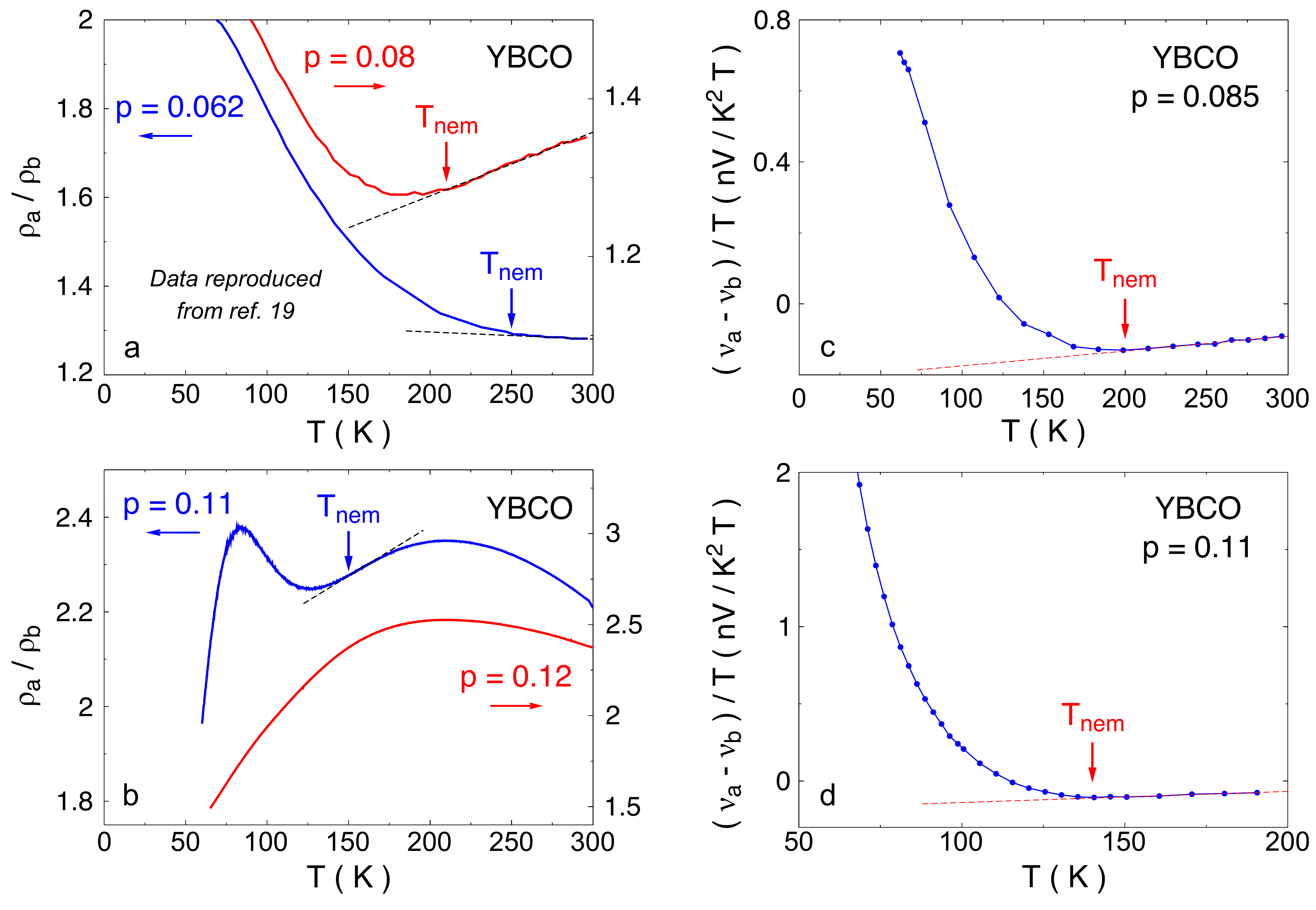}
\caption{(a, b) Temperature dependence of the anisotropy in the in-plane resistivity of YBCO above $T_{\rm c}$, measured as the ratio of 
	$\rho_a$ to $\rho_b$, the electrical resistivities perpendicular ($\rho_a$) and parallel ($\rho_b$) to the CuO chains of the orthorhombic structure. 
	We define the onset temperature for nematicity, $T_{\rm nem}$ (down-pointing arrows), as the start of the upturn as temperature is reduced. 
	$T_{\rm nem}$ is then plotted in the phase diagram of Fig.~\ref{Phasediagram}. 
	(a) Zoom on the data of ref.~\onlinecite{Ando2002}, for two dopings as indicated. 
	(b) Anisotropy ratio $\rho_a\,/\,\rho_b$ in our samples, at two dopings as indicated. 
	At $p = 0.11$ (blue), an upturn is clearly seen, below $T_{\rm nem} = 150 \pm 20$\,K (down-pointing arrow). 
	At $p = 0.12$ (red), no upturn is detected. 
	In the $p = 0.11$ curve, the drop in $\rho_a\,/\,\rho_b$ below 80\,K coincides with the drop in the Hall coefficient 
	at $p = 0.11$ (ref.~\onlinecite{LeBoeuf2011}), due to Fermi-surface reconstruction. 
	(c,~d)~In-plane anisotropy difference in the Nernst coefficient $\nu_x(T)$, plotted as $(\,\nu_a - \nu_b\,)\, /\,T$ (blue dots), 
	for $p = 0.085$ ($H=16$\,T; c) and $p = 0.11$ ($H = 18$\,T; d). 
	The data shown here are independent of the magnetic field strength, $H$. The red dashed lines are a fit to the linear dependence at high temperature. 
	The red arrows mark the onset of a large upturn as temperature is reduced, at temperatures that fall on the $T_{\rm nem}$ line of Fig.~\ref{Phasediagram} (red squares).
	}
\label{Anirho-AniNernst}
\end{figure*}

%%%%%%%%%%%%%%%%%%%%%%%%%%%%%%%%%%%%%%%%%%%%%%%%%%%%%%%%%%%%%%%%%%%%%%

In Fig.~\ref{Anirho-AniNernst}(a), we reproduce data from Ando \etal{}~[\onlinecite{Ando2002}] on YBCO at two dopings, 
and define the temperature $T_{\rm nem}$ as the onset of this upturn. 
We find $T_{\rm nem} = 250 \pm 20$\,K and $210 \pm 20$\,K at $p = 0.062$ and $p = 0.08$, respectively. 
In Fig.~\ref{Phasediagram}, we plot $T_{\rm nem}$ values so obtained on the phase diagram of YBCO. 

In Fig.~\ref{Anirho-AniNernst}(b), we show the anisotropy ratio $\rho_a\,/\,\rho_b$ vs $T$ measured in our own YBCO crystals. 
At $p = 0.11$, we observe a clear upturn, below $T_{\rm nem} = 150 \pm 20$\,K . 
In Fig.~\ref{Phasediagram}, we see that this value is in smooth continuation of the $T_{\rm nem}$ line at lower doping. 
Close inspection of earlier data~[\onlinecite{Ando2002}] at $y = 6.70$ ($p = 0.104$) reveals a faint anomaly at $T \sim 160$\,K, 
which our oxygen-ordered samples bring out more clearly. 
With $T_{\rm nem} \sim T_{\rm XRD}$ at $p = 0.11$, we discover that the nematic line hits the CDW dome near its peak (Fig.~\ref{Phasediagram}). 
Going up to $p = 0.12$, we see no sign of any upturn in $\rho_a\,/\,\rho_b$ (Fig.~\ref{Anirho-AniNernst}(b)), in agreement with previous work~[\onlinecite{Daou2010,Ando2002}] 
(Fig.~\ref{Anirho-Aninu-p08-p12}(b)). 
%However, there could be a downturn below $\sim 100$\,K, as clearly seen below $\sim 80$\,K for $p = 0.11$ (Fig.~\ref{Anirho-AniNernst}(b)). 

In order to detect $T_{\rm nem}$ in a second observable, we measured the Nernst coefficient $\nu_x(T)$ in YBCO at $p = 0.085$ and $p = 0.11$. 
In Figs.~\ref{Anirho-AniNernst}(c) and \ref{Anirho-AniNernst}(d) we plot the in-plane anisotropy difference, as $(\,\nu_a - \nu_b\,)\, /\,T$ vs $T$. 
(Note that our data at $p = 0.085$ are in good agreement with data at $p=0.078$ from ref.~\onlinecite{Daou2010}.) 
With decreasing temperature, we can readily identify the onset of an upturn at $200 \pm 20$\,K and $140 \pm 20$\,K for $p = 0.085$ and $p = 0.11$, respectively. 
These onset temperatures are in excellent agreement with the $T_{\rm nem}$ line (Fig.~\ref{Phasediagram}), 
thereby confirming the location of this line with a thermo-electric measurement.

In Fig.~\ref{Dnu-Tnu-TXRD}, we compare the Nernst anisotropy at five dopings, from $p = 0.085$ to $p = 0.15$. 
We see that the curves split into two groups: those associated with CDW modulations, at $p = 0.12$ and above, whose upturn is correlated with $T_{\rm XRD}$, 
and those associated with the second kind of nematicity, at $p = 0.11$ and below, which do not collapse onto the common curve 
(whether scaled by $T_{\rm XRD}$ or $T^{\star}$), but start their upturn at $T_{\rm nem}$. 
This shows that the upturn in $(\,\nu_a - \nu_b\,)\,/\,T$ is not correlated with $T_{\rm XRD}$, or with $T^{\star}$, across all dopings.
We see that there are in fact two types of nematicity in YBCO, with distinct characteristic temperatures in the phase diagram (Fig.~\ref{Phasediagram}): 
one associated with $T_{\rm nem}$ -- seen in both $\nu$ and $\rho$ -- and another associated with $T_{\rm XRD}$ -- seen only in $\nu$.
In the remainder, we focus on the former nematicity, at $p \leq 0.11$.

%%%%%%%%%%%%%%%%%%%%%%%%%%%%%%%%%%%%%%%%%%%%%%%%%%%%%%%%%%%%%%%%%%%%%%%

\begin{figure}[t]
\centering
\includegraphics[width=0.46\textwidth]{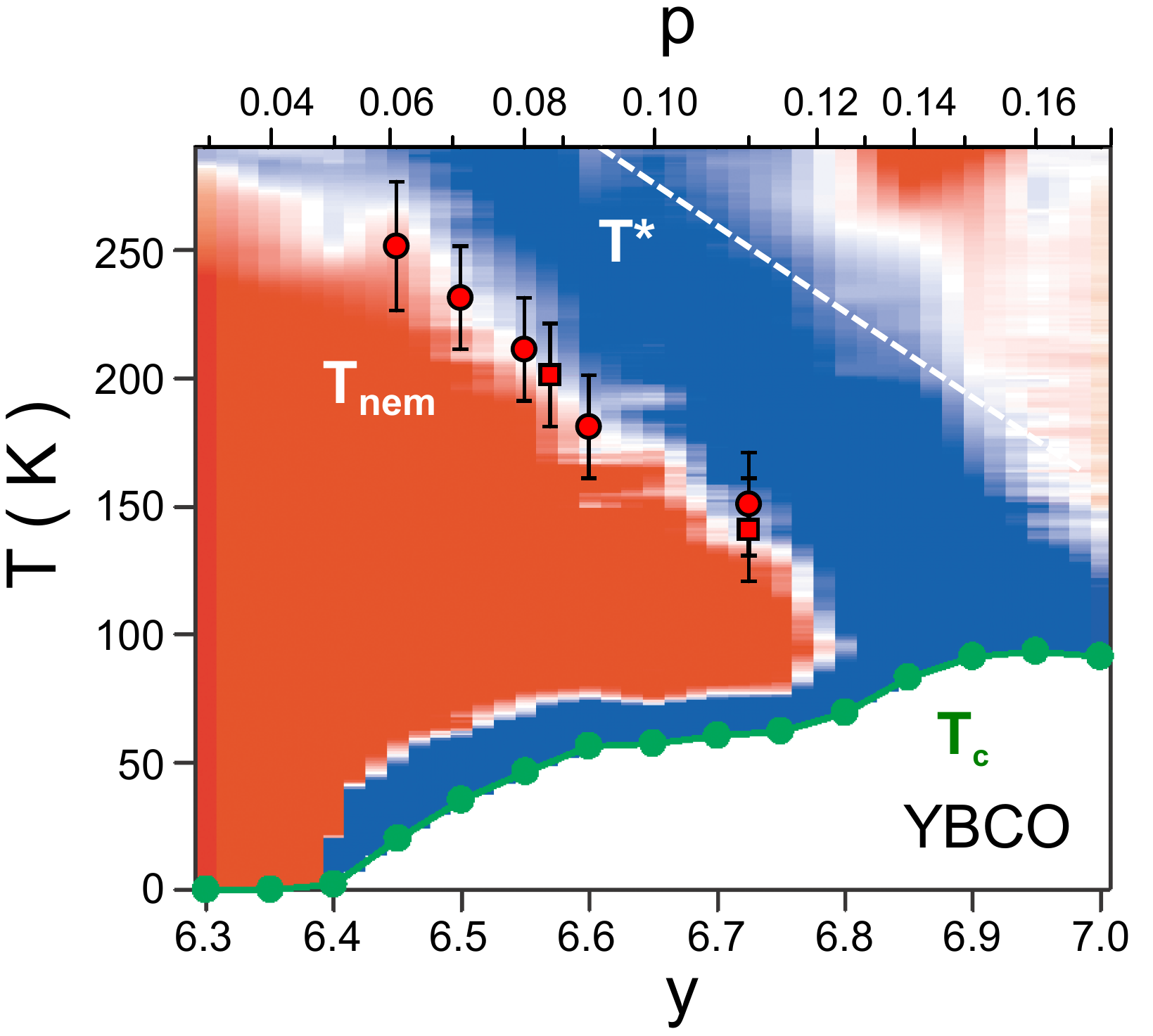}
\caption{Map of the curvature in the resistivity of YBCO, showing the second temperature derivative of $\rho_a(T)$ in the $T-y$ plane, 
	where $y$ is the oxygen content (from ref.~\onlinecite{Ando2004}). 
	Red is for positive curvature, blue for negative curvature, and white for no curvature. 
	The superconducting transition temperature $T_{\rm c}$ is shown as solid green circles, with the corresponding $p$ values given on the top. 
	The onset of nematicity at $T_{\rm nem}$ is shown as red symbols (from Fig.~\ref{Phasediagram}). 
	$T_{\rm nem}$ is seen to lie right at the inflexion point in the resistivity, $T_{\rm x}$, where $\rho_a(T)$ goes from downward (blue) to upward (red) curvature. 
	The lower boundary of the white region of linear-$T$ resistivity at high temperature (upper right corner) 
	is traditionally used to define the pseudogap temperature $T^{\star}$. The white dashed line is the $T^{\star}$ line obtained from Nernst data~[\onlinecite{Daou2010}]. 
	(The blue band above $T_{\rm c}$, of width 20-30\,K, is due to paraconductivity, \textit{i.e.} superconducting fluctuations.)
	}
\label{Ando-Curvature-map}
\end{figure}

It is instructive to overlay our $T_{\rm nem}$ data on the curvature map produced by Ando~\etal~[\onlinecite{Ando2004}], 
where they plot the second derivative of $\rho_a(T)$ throughout the $T-p$ plane. This is done in Fig.~\ref{Ando-Curvature-map}. 
Upon cooling from room temperature, $\rho_a(T)$ in YBCO deviates downward from its linear-$T$ dependence at high $T$ 
below a temperature labeled $T^{\star}$ $-$ the upper boundary of the blue region of negative curvature. 
Traditionally called the ``pseudogap temperature'', $T^{\star}$ is also detected as a drop in $\nu_x\, /\,T$ (dashed line; from ref.~\onlinecite{Daou2010}). 
It is roughly the temperature below which the NMR Knight shift $K_{\rm s}(T)$ starts to drop in YBCO~[\onlinecite{Alloul1989}], for example.

%%%%%%%%%%%%%%%%%   Figure  Ando-Curvature-map  %%%%%%%%%%%%%%%%%%%%%%%%%%%%%%%%%%%%%%%%%%%%%%%%%

%%%%%%%%%%%%%%%%%%%%%%%%%%%%%%%%%%%%%%%%%%%%%%%%%%%%%%%%%%%%%%%%%%%%%%

As $\rho_a(T)$ decreases below $T^{\star}$, it goes through a well-defined inflexion point, at $T_{\rm x}$ $-$ the white line in Fig.~\ref{Ando-Curvature-map}
between 
the blue region of negative curvature at intermediate temperatures and the red region of positive curvature at low temperature. 
We immediately notice that the $T_{\rm nem}$ data points coincide perfectly with the inflexion line in $\rho_a(T)$, \textit{i.e.} $T_{\rm nem} = T_{\rm x}$, at all dopings. 
This comes from the fact that the upturn in the anisotropy ratio $\rho_a\,/\,\rho_b$ is due to the upward curvature in $\rho_a(T)$ itself, which starts at $T_{\rm x}$. 
This reinforces the fact that there exists a real line in the phase diagram that is distinct from the CDW dome and separate from $T^{\star}$. 

Note that nematicity was also observed in \lsco{} (LSCO) at low doping, with comparable values of $T_{\rm nem}$~[\onlinecite{Ando2002}], 
and the region of upward curvature in the curvature map of LSCO has a boundary similar to that of YBCO~[\onlinecite{Ando2004}]. 
Moreover, the in-plane resistivity of the cuprate \hbco~(Hg-1201) has an inflexion point very similar to that found in YBCO at $T_{\rm x}$~[\onlinecite{Barisic2008}]. 
It has been emphasized that $\rho_a(T)$ has a Fermi-liquid-like $T^2$ dependence below a temperature $T^{\star\star} \simeq T_{\rm x}$
in both YBCO and Hg-1201~[\onlinecite{Barisic2013}]. 
From these similarities, we infer that the onset of nematicity at $T_{\rm nem} = T_{\rm x}$ is likely to be a generic property of hole-doped cuprates.

%%%%%%%%%%%%%%%%%%%%%%%%%%%%%%%%%%%%%%%%%%%%%
%%%%%%%%%%%%%                  DISCUSSION                     %%%%%%%%%%%%%
%%%%%%%%%%%%%%%%%%%%%%%%%%%%%%%%%%%%%%%%%%%%%

\section{DISCUSSION}

\begin{figure}[t]
\centering
\includegraphics[width=0.46\textwidth]{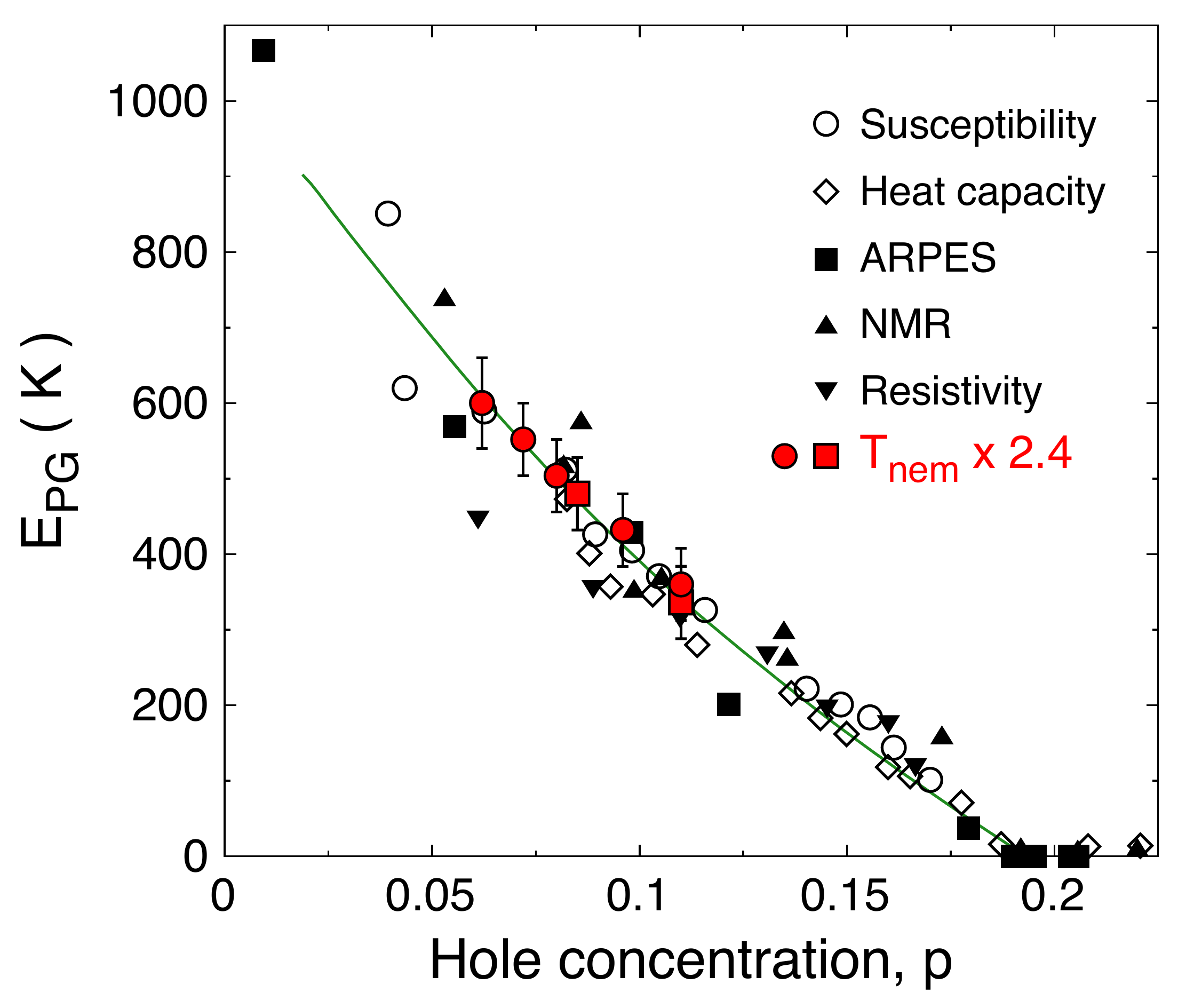}
\caption{Doping dependence of the energy gap $E_{\rm PG}$ extracted from five normal-state properties above $T_{\rm c}$, as indicated 
   (adapted from ref.~\onlinecite{Tallon2001}). 
	All data were obtained on YBCO except for ARPES data, measured on Bi-2212~[\onlinecite{Tallon2001}]. 
	$E_{\rm PG}$ is the pseudogap energy, expressed in K. 
	We compare $E_{\rm PG}$ with our temperature $T_{\rm nem}\,(=T_{\rm x})$ (red symbols, from Fig.~\ref{Phasediagram}), multiplied by a constant factor 2.4. 
	We observe that $T_{\rm nem} \sim E_{\rm PG}$, strong evidence that $T_{\rm nem}$ is a characteristic temperature of the pseudogap phase. 
	}
\label{EPG-Tnem}
\end{figure}

To summarize, two distinct features of $\rho(T)$ $-$ inflexion point and nematicity $-$ are intimately linked, and remain linked at all dopings, 
along a line that decreases linearly with $p$, the $T_{\rm nem} = T_{\rm x}$ line (Figs.~\ref{Phasediagram} and \ref{Ando-Curvature-map}). 
This $T_{\rm nem}$ line has two important properties. The first is that it scales with the pseudogap energy. 
Indeed, as shown in Fig.~\ref{EPG-Tnem}, $T_{\rm nem} \sim E_{\rm PG}$, where $E_{\rm PG}$ is the energy gap extracted from an analysis 
of several normal-state properties of YBCO (and other cuprates) above $T_{\rm c}$, including $K_{\rm s}(T)$, $\rho_a(T)$ and the specific heat~[\onlinecite{Tallon2001}]. 
This is consistent with analyses showing that $K_{\rm s}(T)$ and $\rho_a(T)$ can be scaled on top of each other at all dopings, 
when plotted vs $T\,/\,T_{\rm x}$~[\onlinecite{Wuyts1996,Alloul2014}]. 
The simultaneous rise of $T_{\rm nem}$ and $E_{\rm PG}$ with decreasing $p$ in YBCO (Fig.~\ref{EPG-Tnem}) 
is reminiscent of the simultaneous onset of nematicity and 
pseudogap in \bscco~(Bi-2212) detected in low-temperature STM measurements when doping is reduced below $p \simeq 0.19$~[\onlinecite{Fujita2014}].

%%%%%%%%%%%%%%%%%   Figure  EPG-Tnem  %%%%%%%%%%%%%%%%%%%%%%%%%%%%%%%%%%%%%%%%%%%%%%%%%

%%%%%%%%%%%%%%%%%%%%%%%%%%%%%%%%%%%%%%%%%%%%%%%%%%%%%%%%%%%%%%%%%%%%%%

The second important property of the $T_{\rm nem}$ line is that it hits the $T_{\rm XRD}$ dome at its peak (Fig.~\ref{Phasediagram}). 
This suggests a link between nematicity onset and CDW dome. 
It points to a nematic phase that competes with CDW order, providing a potential explanation for why CDW order and modulations peak at $p = 0.12$. 

Let us consider two possible scenarios for the nematicity that develops in YBCO below $T_{\rm nem}$. 
The first scenario is a nematic phase that breaks the rotational symmetry of the CuO$_2$ planes~[\onlinecite{Kivelson1998,Nie2014}]. 
Calculations of transport anisotropies applied to YBCO showed that the sign and magnitude of the measured $\sigma_b\,/\,\sigma_a$ and $\nu_b\,/\,\nu_a$ 
are natural consequences of such electron nematic order~[\onlinecite{Hackl2009}]. 
Here $T_{\rm nem}$ would correspond to the onset of anisotropic spin correlations, observed with neutrons~[\onlinecite{Hinkov2008,Haug2010}]. 
At low temperature, these correlations turn into short-range spin-density-wave (SDW) order~[\onlinecite{Haug2010}]. 
The SDW order is known to compete with both superconductivity and CDW order~[\onlinecite{Blanco-Canosa2013}], 
hence it may well be responsible for the fall of both $T_{\rm c}$ and $T_{\rm XRD}$ below $p \sim 0.12$ (Fig.~\ref{Phasediagram}). 
This is analogous to a CDW scenario in which $T_{\rm XRD}$ marks the onset of ``charge nematicity"~[\onlinecite{Wu2015}], 
and the onset of long-range CDW order occurs only at the lower temperature $T_{\rm NMR}$~[\onlinecite{Wu2011,Wu2013}]. 
Recent calculations of the in-plane anisotropy of resistivity show how SDW fluctuations can give rise to an upturn in $\rho_a\,/\,\rho_b$, 
while CDW modulations would produce a downturn~[\onlinecite{Schutt2015}]. 
This could explain the two distinct behaviors seen in the resistivity anisotropy of YBCO for $p \leq 0.11$ and $p \geq 0.12$. 

In the second scenario, $T_{\rm nem}$ marks the onset of an enhanced nematic susceptibility, with no broken rotational symmetry. 
In solutions of the Hubbard model, it has been shown that in the presence of a small orthorhombic distortion, of the kind present in YBCO, 
a large in-plane anisotropy of the resistivity appears in the pseudogap phase~[\onlinecite{Okamoto2010,Su2011}]. 
This could explain the upturns seen in the transport anisotropy of YBCO at low $p$ and low $T$, and the correlation between $T_{\rm nem}$ and $E_{\rm PG}$. 
In these calculations, the key organizing principle of the pseudogap phase is the Widom line, a line in the $T-p$ phase diagram where 
electronic properties of the material are predicted to have a distinctive anomaly~[\onlinecite{Sordi2012}]. 
For example, the charge compressibility peaks at the Widom line and both the density of states (\textit{e.g.} the pseudogap measured in specific heat) 
and the spin susceptibility (\textit{i.e.} the Knight shift measured in NMR) go through an inflexion point~[\onlinecite{Sordi2012}]. 
%Note that the gradual onset of the drop in these quantities is at the higher temperature $T^{\star}$~[\onlinecite{Sordi2013}]. 
Given that the onset of nematicity in YBCO at $T_{\rm nem}$ coincides with the inflexion point in $\rho_a(T)$ (at $T_{\rm x}$),
it is tempting to identify the empirical $T_{\rm nem} = T_{\rm x}$ line of Fig.~1 with the theoretical Widom line of ref.~\onlinecite{Sordi2012}.

%%%%%%%%%%%%%%%%%%%%%%%%%%%%%%%%%%%%%%%%%%%%%
%%%%%%%%%%%%%                  SUMMARY                     %%%%%%%%%%%%%%
%%%%%%%%%%%%%%%%%%%%%%%%%%%%%%%%%%%%%%%%%%%%%

\section{SUMMARY}

Nematicity in YBCO is detected as the spontaneous appearance of an in-plane anisotropy
in the transport properties upon cooling, in addition to the background anisotropy due to the CuO chains
that run along the $b$ axis of the orthorhombic structure.
Two types of nematicity exist in two different regions of the temperature-doping phase diagram (Fig.~1).
At low doping ($p \leq 0.11$), the first type is detected in both the resistivity and the Nernst coefficient,
as parallel upturns in the anisotropy ratios, $\rho_a / \rho_b$ and $\nu_b / \nu_a$ (Fig.~\ref{Anirho-Aninu-p08-p12}).
At high doping ($p \geq 0.12$), the second type is detected only in the Nernst coefficient,
and not in the resistivity (Fig.~\ref{Anirho-Aninu-p08-p12}).
We attribute this second nematicity to the short-range CDW modulations detected in YBCO by X-ray diffraction,
which grow gradually upon cooling below $T_{\rm XRD}$ (Fig.~1), since 
the anisotropy ratio $\nu_b / \nu_a$ at $p \geq 0.12$ grows in tandem with the CDW intensity (Fig.~\ref{Dnu-RX}).

We discuss two possible scenarios for the first type of nematicity ($T_{\rm nem}$~line in Fig.~1).
In the first scenario, rotational symmetry is broken below $T_{\rm nem}$ as a precursor to the 
onset of SDW modulations at lower temperature, which then also break translational symmetry.
In the second scenario, there is no broken rotational symmetry but only an enhanced nematic
susceptibility appearing gradually below $T_{\rm nem} = T_{\rm x} $ (Fig.~\ref{Ando-Curvature-map}), due to the emergence of the pseudogap phase.
This could explain why $T_{\rm nem}$ scales with the pseudogap energy $E_{\rm PG}$ (Fig.~\ref{EPG-Tnem}).

%%%%%%%%%%%%%%%%%%%%%%%%%%%%%%%%%%%%%%%%%%%%%%%%%%%%
%%%%%%%%%%%%%                  ACKNOWLEDGEMENTS                     %%%%%%%%%%%%%%
%%%%%%%%%%%%%%%%%%%%%%%%%%%%%%%%%%%%%%%%%%%%%%%%%%%%

\section*{ACKNOWLEDGEMENTS }
We would like to thank S. A. Kivelson, E. Fradkin, A.-M. Tremblay, G. Sordi, D. S\'{e}n\'{e}chal, S. Sachdev, 
J. Chang, C. Proust, A. Georges, D. LeBoeuf, C. P\'{e}pin, R. Fernandes, A. Millis, M. Sch\"{u}tt, 
K. Behnia, H. Alloul, P. Bourges, M.-H. Julien, R. Greene, J. Tallon, I. Fisher, M. Le Tacon, A. Damascelli, 
O. Parcollet, M. Ferrero, B. Keimer, Y. Ando, A. V. Chubukov, M. Vojta, 
S. Dufour-Beaus\'{e}jour and F. F. Tafti 
for fruitful discussions and
M. Caouette Mansour, J. Corbin and S. Fortier for their assistance with the experiments at Sherbrooke. 
L.T. thanks ESPCI-ParisTech, Universit\'{e} Paris-Sud, CEA-Saclay and the Coll\`{e}ge de France for their hospitality and support, 
and the \'{E}cole Polytechnique (ERC-319286 QMAC) and LABEX PALM (ANR-10-LABX-0039-PALM) for their support, while this article was written. 
R.L., D.A.B. and W.N.H. acknowledge funding from the Natural Sciences and Engineering Research Council of Canada (NSERC). 
L.T. acknowledges support from the Canadian Institute for Advanced Research (CIFAR) 
and funding from NSERC, the Fonds de recherche du Qu\'{e}bec - Nature et Technologies (FRQNT), the Canada Foundation for Innovation, and a Canada Research Chair.

%%%%%%%%%%%%%%%%%%%%%%%%%%%%%%%%%%%%%%%%%%%%%%%%%%%%
%%%%%%%%%%%%%                  BIBLIOGRAPHY                          %%%%%%%%%%%%%%
%%%%%%%%%%%%%%%%%%%%%%%%%%%%%%%%%%%%%%%%%%%%%%%%%%%%

%

%%%%%%%%%%%%%%%%%%%%%%%%%%%%%%%%%%%%%%%%%%%%%%%%%%%%%%%%%%

%%%%%%%%%%%%%%%%%%%%%%%%%%%%%%%%%%%%%%%%%%%%%%%%%%%%%%%%%%

\end{document}